\documentclass[a4paper]{article}

\usepackage[affil-it]{authblk}
\usepackage{natbib}
\usepackage{hyperref}
\usepackage{amsmath}
\usepackage{xcolor}
\newcommand{\code}[1]{\texttt{#1}}
\usepackage[section]{placeins}

\usepackage{graphicx}

\begin{document}

\title{Directional genetic differentiation and relative migration}

\author{\small Lisa Sundqvist}
 \affil{\small Corresponding author. University of Gothenburg, Department of Marine Sciences, SE-405 30 Gothenburg, Sweden, lisa.sundqvist@gu.se, Fax:+46317862560}

\author{Kevin Keenan}
  \affil{Queen's University Belfast, Institute for Global Food 
    Security, School of Biological Sciences, Belfast BT9 7BL, Northern Ireland, 
    United Kingdom}
  
\author{Martin Zackrisson}
  \affil{University of Gothenburg, Department for Chemistry
    and Molecular Biology, SE-405 30 Gothenburg, Sweden.}
    
    \author{Paulo Prod{\"o}hl}
  \affil{Queen's University Belfast, Institute for Global Food 
    Security, School of Biological Sciences, Belfast BT9 7BL, Northern Ireland, 
    United Kingdom}
    
    \author{David Kleinhans}
  \affil{Carl von Ossietzky University, ForWind Center for Wind Energy Research, Institute of
    Physics, DE-26129 Oldenburg, Germany.}

\maketitle

{\small \textbf{Keywords:} Directional gene flow, Asymmetric migration, Dispersal, Allele frequency data. Running title: Directional relative migration}

\begin{abstract}

Understanding the population structure and patterns of gene flow within species is of fundamental importance to the study of evolution. In the fields of population and evolutionary genetics, measures of genetic differentiation are commonly used to gather this information. One potential caveat is that these measures assume gene flow to be symmetric. However, asymmetric gene flow is common in nature, especially in systems driven by physical processes such as wind or water currents. Since information about levels of asymmetric gene flow among populations is essential for the correct interpretation of the distribution of contemporary genetic diversity within species, this should not be overlooked. To obtain information on asymmetric migration patterns from genetic data, complex models based on maximum likelihood or Bayesian approaches generally need to be employed, often at great computational cost. Here, a new simpler and more efficient approach for understanding gene flow patterns is presented. This approach allows the estimation of directional components of genetic divergence between pairs of populations at low computational effort, using any of the classical or modern measures of genetic differentiation. These directional measures of genetic differentiation can further be used to calculate directional relative migration and to detect asymmetries in gene flow patterns. This can be done in a user-friendly web application called divMigrate-online introduced in this paper. Using simulated data sets with known gene flow regimes, we demonstrate that the method is capable of resolving complex migration patterns under a range of study designs. 
 
  \end{abstract}
  

\section{\label{sec:introduction}Introduction}

Measures of population genetic differentiation are widely used in studies 
focusing on conservation, management and evolution.
Generally, information about the pattern of population 
structure and the level of differentiation are derived from allele frequency data.
The most commonly used measures, utilizing this information, being Wright's 
fixation index $F_{st}$ \citep{Wright43}, Nei's $G_{st}$ \citep{Nei73}, 
and the more recently introduced  $G'_{st}$ \citep{hedrick2005} and $D$\citep{Jost08}, which are independent of gene diversity.

A particularly useful feature of these parameters is that, assuming an island 
model of population structure, they can be used to estimate migration among 
populations \citep{Wright31, Wright49, Jost08}. These measures, however, assumes
migration to be symmetric (i.e. of equal rate in all directions). 
For simplification, the term migration is here and for the rest of this paper 
used interchangeably with the term gene flow. 

Structured populations characterized by asymmetric migration is, on the other hand, common 
in nature. Especially in systems driven by physical transport processes, such as 
wind or water currents \citep{Pringle11}, or those with strong habitat 
quality gradients, which can lead to competition driven directional dispersal 
\citep{paz-vinas2013}. In marine environments, ocean currents are known to affect both the 
dispersal of planktonic and benthic species, the latter often being characterized with a planktonic phase 
during early stages of development \citep{Siegel03,Cowen09}. 
Other examples, where the direction of gene flow can be affected by 
physical processes,  include organism living in river systems (e.g. 
impassable waterfalls prevent upstream dispersal), as well as wind 
pollinated plants, mosses and lichen \citep{Munoz04, Hanfling06, Friedman09}.

Both physical processes and density dependent competition can lead to 
source-sink dynamics among populations (i.e. metapopulation structure).
Where this occurs, parameters such as genetic differentiation can 
be significantly skewed \citep{Dias96}. This, in turn, can lead to incorrect conservation and/or management decisions. For instance, while sink populations of 
poor quality (i.e. low/no intra-population recruitment) contribute little to 
the long-term evolutionary potential of the species \citep {Whitlock99} they can display greater genetic differentiation relative to source populations \citep {Whitlock99}. As a consequence, they can be incorrectly
regarded as a 'unique' component of the species' overall diversity, 
which may misdirect conservation efforts. 

In asymmetric systems, it becomes important to estimate directional migration
to fully understand the processes leading to genetic structuring of 
populations, as well as to allow for effective conservation and management 
decisions to be derived from such information \citep[e.g.][]{Pringle11}.
For example, when aiming to conserve endangered species, implementing
management strategies for source populations is known to be more effective
than for sink populations \citep{Dias96}.
In addition to the difficulties associated with understanding genetic structure,
the absence of information about the symmetry of gene flow can also make
the inference of past demographic changes within populations more difficult.
For instance, \citet{paz-vinas2013} recently demonstrated,
using both simulated and empirical data, that the probability of incorrectly 
inferring past population expansions increased dramatically when populations 
were characterized by asymmetric patterns of gene flow.
Thus, the ability to detect asymmetric migration, when it is present, is
essential for the understanding of evolutionary processes that have led to contemporary
diversity patterns.

Currently, to obtain information about patterns of migration from genetic 
data in asymmetric systems, complex mathematical models using maximum 
likelihood or Bayesian approaches are typically required 
\citep[e.g.][]{Wilson03, Beerli09}. In these models, a large number of parameters are estimated simultaneously involving complex optimization algorithms, resulting in intensive 
computational requirements. As a consequence, these models are often used as \emph{black boxes} implying 
that users, due to the complexity of these analytical approaches, typically only have a 
limited understanding of the underlying models and their assumptions. Thus, most users are not in a position to adequately assess when their application 
is inappropriate.

In contrast, the present study introduces a new, relatively simple and 
tangible method for the detection of asymmetric migration from allele 
frequency data. This approach provides robust information on the direction 
of migration and is intended to fill the gap between existing complex methods 
for measuring asymmetric migration and symmetric measures of genetic differentiation.
The method is based on defining a \emph{hypothetical pool of migrants}
for a given pair of populations and estimating an appropriate 
measure of genetic differentiation between each of the two populations and 
the hypothetical pool. The directional genetic differentiation can then be used to estimate the relative levels of migration between the two populations.
The larger of the two relative migration values indicates the population 
that is likely to behave as a source population (i.e. the hypothetical pool of migrants is 
genetically more similar to this population), while the smaller of the two 
estimates indicates the population most likely to behave as a sink.
By testing whether these two genetic distances are significantly
different from one another it is also possible to determine 
whether migration occurs at a significantly higher rate in one direction over 
the other. The concept of a hypothetical pool of migrants makes it possible to gain new information about direction of gene flow using regular symmetric measures of genetic differentiation.

This new approach opens up a new dimension of applications where directional
measures of genetic differentiation can be used both to explore and detect 
asymmetric migration patterns.
It is argued that this tool will be of major utility in molecular ecological studies where 
information about the presence/absence of genetic connectivity and its 
dynamics among populations is of interest. Consider, for example, the 
benefit of being able to explicitly identify source and sink populations 
in metapopulations; having the ability to test whether physical barriers 
precluded gene flow in a particular direction; or the ability to understand the 
influence of ecosystem processes on gene flow (e.g. is the direction of migration
among populations correlated with an ocean current).
Accordingly, we present this method as a new tool to 
researchers interested in a better understanding of the demographic and 
evolutionary dynamics of populations and species and more specifically on the use of this 
information for conservation and management.


\section{\label{sec:theory}Theory}

\subsection{\label{sec:meas-genet-diff}Measures of genetic
  differentiation}

For measures of genetic differentiation a value of zero
indicate that the allele frequencies among populations are equal, while
values larger then zero represent increasing differences \citep{Meirmans11}. 
Generally, these genetic differentiation measures are based on two parameters, 
which describe the distribution of genetic diversity among populations, 
namely, the mean heterozygosity in the total population ($H_t$) and the mean 
heterozygosity within the individual populations ($H_s$) \citep{Meirmans11}.
While an extensive formal set of notations exists for these two parameters, 
as well as for $G_{st}$ and $D$, in here we introduce vector notations which make it
possible to carry out calculations for a number of populations simultaneously.

Let the total number of different alleles present in $P$ populations
be $N$. For now, equal population sizes are assumed. The allele frequencies in the individual populations can then be arranged in the $N\times P$-matrix $A$.

\begin{equation}
  \label{eq:allel-frequency-matrix}
  A=\left(\begin{array}{llll}a_{11}&\cdots&a_{1P}\\\vdots&\cdots&\vdots\\a_{N1}&\cdots&a_{NP}\end{array}\right)=\left(\boldsymbol{a_1},\cdots,\boldsymbol{a_P}\right)\quad,
\end{equation}
where the matrix element $a_{ij}$ represents the frequency of allele
$i$ in population $j$ with $\sum_ia_{ij}=1$ for any population
$j$.
The column vectors $\boldsymbol{a_j}\in[0,1]^N$ constitute the
allele frequencies in the individual populations, please
  note that bold letters indicate vectors. For each population, the
degree of heterozygosity ($H$) can be estimated from the vector of
allele frequencies $\boldsymbol{a}\in[0,1]^N$ as:

\begin{equation}
  \label{eq:heterozygosity}
  {H}(\boldsymbol{a})=1-\boldsymbol{a}^T\boldsymbol{a}=1-|\boldsymbol{a}|^2\quad.
\end{equation}

For a pairwise comparison involving two populations with allele frequencies
$\boldsymbol{a}\in[0,1]^N$ and $\boldsymbol{b}\in[0,1]^N$,
within-population heterozygosity ($H_s$) and total-population
heterozygosity ($H_t$) can be estimated from equation
(\ref{eq:heterozygosity}) as:

\begin{subequations}
  \label{eq:hs-and-ht}
  \begin{eqnarray}
    \label{eq:ht}
    {H_t}(\boldsymbol{a},\boldsymbol{b})&=&1-\frac{1}{4}\left|\boldsymbol{a}+\boldsymbol{b}\right|^2\quad,\\
    \label{eq:hs}
    {H_s}(\boldsymbol{a},\boldsymbol{b})&=&1-\frac{1}{2}\left(|\boldsymbol{a}|^2+|\boldsymbol{b}|^2\right)\quad.
  \end{eqnarray}
\end{subequations}

From these expressions measures for genetic
differentiation between populations with allele frequencies
$\boldsymbol{a}$ and $\boldsymbol{b}$ can be defined using vector
algebra:
\begin{subequations}
  \label{eq:all-measures-undirected}
  \begin{eqnarray}
    \label{eq:Dst-undirected}
    {D_{st}}(\boldsymbol{a},\boldsymbol{b}) &=&
    {H_t}(\boldsymbol{a},\boldsymbol{b})-{H_s}(\boldsymbol{a},\boldsymbol{b})=\frac{1}{4}|\boldsymbol{a}-\boldsymbol{b}|^2\quad,\\
    \label{eq:Gst-undirected}
    {G_{st}}(\boldsymbol{a},\boldsymbol{b})&=&\frac{{D_{st}}(\boldsymbol{a},\boldsymbol{b})}{{H_t}(\boldsymbol{a},\boldsymbol{b})}=\frac{|\boldsymbol{a}-\boldsymbol{b}|^2}{4-|\boldsymbol{a}+\boldsymbol{b}|^2}\quad,\\
    \label{eq:D-undirected}
    {D}(\boldsymbol{a},\boldsymbol{b})&=&\frac{2{D_{st}}(\boldsymbol{a},\boldsymbol{b})}{1-{H_s}(\boldsymbol{a},\boldsymbol{b})}=\frac{|\boldsymbol{a}-\boldsymbol{b}|^2}{|\boldsymbol{a}|^2+|\boldsymbol{b}|^2}\quad.
  \end{eqnarray}
\end{subequations}

Expressions (\ref{eq:all-measures-undirected}) are equivalent to the
(biased) estimators used for calculation of genetic differences as
compiled e.g.\ in \citep{Jost08}.
Multi-locus $D$ is calculated as by \citep{crawford10} and multi-locus
$G_{ST}$ is calculated as proposed in \citep{Nei73}.


\subsection{\label{sec:new-conc-estim} A new concept for estimating
  directional measures of genetic differentiation}

To introduce the new analytic approach, the following hypothetical scenario 
is considered. Assuming two populations A and B with strong directional gene flow in the 
direction A $\rightarrow$ B, but no gene flow in the opposite direction, 
(e.g.\ as observed when waterfalls prevent or restrict movement in a single 
direction for species in river systems).
In addition, population B may exchange genes with other populations that 
are genetically distinct from A and which contain alleles not present in A.
How would such a situation be reflected in the allelic frequencies?
Generally, it can be expected that most alleles present in A are also present 
in population B, whereas alleles present in B may or may not be present in A 
due to the absence of gene flow from B to A. 

In the case of neutral loci the relative allele frequencies of A are expected to be reflected in the migration and therefore 
the proportions of the allele frequencies is assumed to be equal in population B.
An idealized example of an allelic matrix (A) is listed in Table 
\ref{tab:thought-exp}.
In this example alleles 1 and 2, present in population A are represented at 
reduced frequencies but equal proportions in population B, whereas allele 3 is only present in population B. 
From the distribution of allele frequencies it becomes
evident, that there is no gene flow from B to A, but at least some degree of gene flow from A to B cannot be ruled out. 

This concept can be formalized as follows; for each mutual pair of populations to 
be investigated (populations A and B in this example), a hypothetical 
pool of migrants with an allelic composition inferred from the two 
populations surveyed is introduced. That is, the hypothetical population 
has an allelic distribution $\boldsymbol{f}(\boldsymbol{a},\boldsymbol{b})$.
The allelic frequencies represented in the pool are calculated as the (normalized)  geometrical means of the allelic frequencies in the corresponding two populations of interest: $f_i(\boldsymbol{a},\boldsymbol{b})=\gamma \sqrt{a_ib_i}$ with $\gamma=\left(\sum_i \sqrt{a_ib_i}\right)^{-1}$ if $\sum_i \sqrt{a_ib_i}>0$ and $\gamma=0$ otherwise. 

Let us present a general motivation of this definition of the hypothetical pool of migrants. From the hypothetical scenario a number of requirements for the hypothetical pool
$\boldsymbol{f}(\boldsymbol{a},\boldsymbol{b})$ are defined:
\begin{enumerate}
\item $\boldsymbol{f}(\boldsymbol{a},\boldsymbol{b})$ should be
  symmetric in its arguments, since the order of populations is
  arbitrary.
\item Alleles not represented in one of the populations can be assumed
  not to be relevant for gene flow.
  As a consequence $\boldsymbol{f}$
  should be non-zero only for alleles present at non-zero frequencies
  in both populations.
\end{enumerate}
A general form for $\boldsymbol{f}(\boldsymbol{a},\boldsymbol{b})$
consistent with these two conditions would be
\begin{equation}
  \label{eq:f-general-multiplicative}
  f_i(\boldsymbol{a},\boldsymbol{b})=\gamma \sum_{k=1}^K\gamma_k\left(a_i^{\alpha_k}b_i^{\beta_k}+a_i^{\beta_k}b_i^{\alpha_k}\right)\quad\forall i
\end{equation}
with arbitrary exponents $\alpha_k,\beta_k>0$ for all $k$, an
arbitrary $K$ defining the number of summands involved and arbitrary weighting and normalization constants $\gamma$ and $\gamma_k$. Here \emph{arbitrary} means, that at this stage any function defined through any arbitrary set of the parameters above would be consistent with considerations 1 and 2. In this vein equation \eqref{eq:f-general-multiplicative} defines a quite general class of functions.

In the optimal example data discussed in connection
with the hypothetical scenario (Table \ref{tab:thought-exp}) 
$a_1/a_2=b_1/b_2$, which mirrors the fact that the proportions of alleles $1$ and $2$ in population 1 is reflected in the gene flow and transferred to population 2.
In order to be consistent with the hypothetical scenario, the pool of
migrants is required to reproduce this proportion of alleles as well:
\begin{enumerate}
  \setcounter{enumi}{2}
\item In case of strong gene flow only occurring in one direction between 
  populations, initially with mutually different alleles, 
  $f_i(\boldsymbol{a},\boldsymbol{b})/f_j(\boldsymbol{a},\boldsymbol{b})=a_i/a_j=b_i/b_j$
  is required for all combination of alleles $i$ and $j$ present in both 
  populations.
\item As a vector of allele frequencies $\boldsymbol{f}$ needs to be
  normalized, i.e.\ it needs to fulfill $\sum_if_i=1$. \end{enumerate}
From the 3rd requirement on $\boldsymbol{f}$ it can be defined that
$\alpha_k+\beta_k=1$ for any $k$, making an equation of the form
\begin{equation}
  \label{f-general-normalized}
  f_i(\boldsymbol{a},\boldsymbol{b})=\gamma\sum_{k=1}^K\gamma_k\left(a_i^{\alpha_k}b_i^{1-\alpha_k}+a_i^{1-\alpha_k}b_i^{\alpha_k}\right)\quad\forall i
\end{equation}
with still arbitrary $0<\alpha_k<1$ for all $k$, the most general 
functional approach for the allelic frequencies in the pooled populations.
In cases where populations do not share any alleles $\boldsymbol{f}$, will not contain any alleles shared with either of these populations. In such cases, genetic differentiation can be set to $1$ in both directions.

Equation \eqref{f-general-normalized} still reflects a vast class of functions, which could meet our conditions. For the time being, and for the rest of this paper, we choose the simplest available 
function in this class, namely
$\alpha_1={\gamma}_1=1/2$ and $K=1$, resulting in
\begin{equation}
  \label{f-geo-mean}
  f_i(\boldsymbol{a},\boldsymbol{b})=\gamma \sqrt{a_ib_i}\quad\forall i
\end{equation}
with $\gamma=\left(\sum_i \sqrt{a_ib_i}\right)^{-1}$.
Hence, the vector of allelic frequencies of the hypothetical population, 
$\boldsymbol{f}(\boldsymbol{a}, \boldsymbol{b})$, is composed of the the normalized
geometrical means of the respective components of $\boldsymbol{a}$ and
$\boldsymbol{b}$.  

As a measure for directional differentiation, both A and B can now be 
compared to their shared hypothetical pool of migrants 
$\boldsymbol{f}(\boldsymbol{a},\boldsymbol{b})$.
For this purpose, any standard, non-directional measures for genetic 
differentiation can be applied. Here we choose to use $G_{st}$ and $D$ as introduced in section \ref{sec:meas-genet-diff}.
 
For the general case of a number of $P$ populations, a
$P\times P$-matrix $B$ can be defined, containing the directional measures for
genetic differentiation, as
\begin{equation}
  B(A,E(\cdot,\cdot))=\left(\begin{array}{llll}b_{11}(A,E(\cdot,\cdot))&\cdots&b_{1P}(A,E(\cdot,\cdot))\\\vdots&\cdots&\vdots\\b_{P1}(A,E(\cdot,\cdot))&\cdots&b_{PP}(A,E(\cdot,\cdot))\end{array}\right)\quad,
\end{equation}
where the individual elements are defined as
\begin{equation}
  b_{ij}(A,E(\cdot,\cdot))=E\left(\boldsymbol{a_i},\boldsymbol{f}\left(\boldsymbol{a_i},\boldsymbol{a_j}\right)\right)\quad\forall
  i,j\quad.
\end{equation}
Here, $E(\cdot,\cdot)$ is a place-holder for estimates of genetic
differentiation, such as $G_{st}$ and $D$. $A$ is the matrix of allele frequencies 
containing the population specific allele frequency vectors $\boldsymbol{a}$.
The value of a particular matrix element $b_{ij}$ can now be taken as an indicator of the
potential for gene flow from population $i$ to population $j$.
Low values indicate a high potential for migration, while high values indicate
low potential for migration in the direction of interest.

With the idealized example data listed in Table \ref{tab:thought-exp}, 
$\boldsymbol{f}(\boldsymbol{a},\boldsymbol{b})=\boldsymbol{a}$, implying that 
the genetic constitution of A is equal to the hypothetical pool of migrants, 
whereas the genetic differentiation between the pool and population B is 
non-zero 
($G_{st}(\boldsymbol{b},\boldsymbol{f}(\boldsymbol{a},\boldsymbol{b}))=0.15$
and
$D(\boldsymbol{b},\boldsymbol{f}(\boldsymbol{a},\boldsymbol{b}))=0.42$).
This result indicates that there is a lower potential for gene flow
from B to A than there is from A to B, which is consistent with the initial 
hypothetical scenario.

It is important to emphasize that, in this paper, we only consider idealized conditions. Thus, the impact of more complex scenarios including, for instance, recent common ancestry, unequal population sizes or populations not being at equilibrium are not considered at this point. The motivation is kept as general and clear as possible to encourage further developments, which allow for adapting the method to more specific and complex conditions. Notwithstanding its simplicity, however, we argue that our model seems to be robust enough to deal with a wide range of conditions. 


\subsection{\label{sec:migration}Estimations of directional relative migration}

Assuming Wright's infinite island model, $F_{st}$ and by extension $G_{st}$ can 
directly be related to migration \citep{Wright31}.

\begin{equation}
  \label{eq:Gst-estimation-formula}
N_{e}m \approx \frac{((\frac{1}{G_{st}}) - 1)}{4}
\end{equation}

Here we use equation \eqref{eq:Gst-estimation-formula} as a step to calculate relative migration, thus either the infinite or finite island model can be used at this point.

Jost outlines in a similar way how $D$ can be used to calculate migration using the finite island model.
If we assume, that the mutation rate does not differ between populations $\mu$ can 
be eliminated from the Equation (22) in Jost (2008) by defining the scaled migration rate $M=m/\mu$. 
We fixed the number of sampled populations $n$ to two, since we only estimate pairwise comparisons and get

\begin{equation}
  \label{eq:jost-relative-formula}
  M\approx (1-D)/D\quad.
\end{equation}

From the effective migration rate in equation \eqref{eq:Gst-estimation-formula}, and the scaled migration rate in equation \eqref{eq:jost-relative-formula}, relative migration rates between population pairs can be calculated. For now we only focus on relative migration rates to make the method less sensitive to possible inaccuracies in the estimation of effective migration rate and scaled migration rate, due to populations not fitting the assumptions of the island model or not being in drift-mutation equilibrium. Extending this concept to the directional measures of genetic differentiation introduced in the previous section, a similar expression can be used for the estimation of directional relative migration. To this end, a 
$P\times P$-matrix $C$ of relative migration is defined as 

\begin{equation}
  C(A)=\left(\begin{array}{llll}c_{11}(A)&\cdots&c_{1P}(A)\\\vdots&\cdots&\vdots\\c_{P1}(A)&\cdots&c_{PP}(A)\end{array}\right)\quad,
\end{equation}

to calculate migration from $G_{st}$  the individual elements are defined as

\begin{equation}
  \label{eq:Gst-estimation-formula2}
c_{ij}(A)=\frac{((\frac{1}{G_{st}\left(\boldsymbol{a_i},\boldsymbol{f}\left(\boldsymbol{a_i},\boldsymbol{a_j}\right)\right)}) - 1)}{4}\quad\forall
  i,j\quad.
\end{equation}

to calculate migration from $D$ the individual elements are defined as 

\begin{equation}
  c_{ij}(A)=\frac{1-D\left(\boldsymbol{a_i},\boldsymbol{f}\left(\boldsymbol{a_i},\boldsymbol{a_j}\right)\right)}{D\left(\boldsymbol{a_i},\boldsymbol{f}\left(\boldsymbol{a_i},\boldsymbol{a_j}\right)\right)}\quad\forall
  i,j\quad.
\end{equation}

The matrix  $C$ is then normalized by its biggest value to obtain directional relative migration ranging from zero to one. $A$ is the matrix of allele frequencies containing the population specific allele 
frequency vectors $\boldsymbol{a}$. The respective matrix elements $c_{ij}(A)$ provide an estimate
for the directional relative migration from population $i$ to population $j$. 
To assess whether gene flow is significantly asymmetric, the directional genetic distances presented in the previous section can be assessed by means of a bootstrap procedure.


\section{Simulations}

\subsection{Methods}

To assess the performance of the method described, we evaluate it under variable gene flow patterns and sample designs. This was done by simulating multiple 
microsatellite data sets from two populations ($N = 1000$), under three distinct gene flow patterns (\emph{unidirectional gene flow}, 
\emph{bidirectional symmetric gene flow} and \emph{bidirectional asymmetric gene flow}). While we acknowledge that these three patterns do not cover all scenarios likely to take place in nature, they are good starting points to begin to assess the potential usefulness of the the new method.

The ability of the proposed new approach to resolve each of these migration 
patterns was tested for three levels of gene flow as follows: low ($m = 0.00025$), medium ($m = 0.005$) and high ($m = 0.05$), 
which correspond to population divergence levels of $F_{st} = 0.5,0.05,0.005$ using the equation $F_{st}\approx\frac1{4N_{e}m+1}$ \citep{Wright31}. 

In cases where the pattern was unidirectional, gene flow was simulated in the 
direction from population two to population one. For the symmetric bidirectional pattern, 
gene flow was simulated in both directions at 
an equal rate ($m/2$). For the asymmetric bidirectional pattern gene flow was simulated in both 
directions, but at $1/4$ from population one to population two and $3/4$ 
from population two to population one ($m*1/4$ and $m*3/4$).

For each combination of gene flow pattern and gene flow rate, 1000 
microsatellite data sets were simulated for each sample size starting with 
$s=10$ and increasing to $s=100$ in increments of 10 
(i.e. $s = 10, 20, ..., 100$). This process was repeated for the 
number of loci, which were also increased from 10 to 100. 
Three gene flow patterns, under three levels of migration, tested 
for 10 different sample sizes and 10 different numbers of loci resulted in 
180 unique simulation scenarios, for each of which 
1000 simulation replicates were generated. When evaluating increasing sample sizes,
the number of loci were fixed at 50 and when evaluating
the number of loci the sample size was fixed to 50.

For each combination of parameters the number of times out of the 1000 replicates that the method detects higher migration from population two to population one is calculated in percent. For the unidirectional and the asymmetric bidirectional gene flow patterns, the method was expected to detect 
higher rate in this direction and the result is expected to approach 100\%.  

In the scenario with symmetric bidirectional gene flow, the method is expected to 
estimate equally high migration rates in both directions, therefore we do not expect to see a signal. In this 
scenario, due to random chance, half of the simulations are expected to be higher in one direction and the other half to be higher in the other direction. Thus a value of 50\% is the expected outcome.

In all simulations, mutation rate was fixed at $ 5 \times 10^{-4}$, a 
value which is thought to be common at microsatellite loci in vertebrates 
\citep{Bhargava10}, and base population size was fixed at $2N = 2000$. 
The performance of the directional migration method was assessed by 
calculating relative migration rates derived from both $D$ and $G_{st}$.
Relative migration rates were calculated using a newly developed \code{R} function, 
\code{divMigrate} from the \code{diveRsity} package \citep{Keenan13}. Simulations were carried out using \code{fastsimcoal2} \citep{Excoffier11}, 
and the whole analysis process was scripted in \code{R}.
Instructions and R scripts to enable readers to reproduce these analyses are 
available in a dedicated github repository (\url{https://github.com/lisasundqvist/Sundqvist_et_al_2016}).

\subsection{Results}

\subsubsection{Unidirectional gene flow}

The ability of the method to detect the simulated migration from population two 
to population one was best for the medium migration rate ($m = 0.005$). When the sample size reached 20, the correct direction was found in 95\% and more of the simulations for both $D$ and $G_{st}$ (figure 1a and 1b). When the number of loci were increased, the result exceeded 95\% when the number of loci was 40 for $D$ (figure 2a) and 20 for $G_{st}$ (figure 2b). When the number of loci were 60 or higher the calculations using $G_{st}$ reached 100\% correct directions (figure 2b), indicating that the correct direction was detected in all of the 1000 simulations.

The number of correct directions estimated by the method was next highest for low level 
of migration. For increasing sample size, calculations from $G_{st}$ reached 90\% correct directions when the sample size reached 40. When number of loci varied the number of correct directions reached 90\% for $D$ when the number of loci was 70 (figure 2a). For $G_{st}$, the result reached 90\% when the number of loci was 50 and 95\% when the number of loci reached 80 (figure 2b).
 
When gene flow was high ($m = 0.05$), the least number of correct directions was estimated coming close and up to 75\% only for the highest sample size and number of loci tested (figure 1 and 2). Increasing sample size had highest effect between 10 and 30 (figure 1). After that, increase in sample size only improve the results when migration was high. 

\subsubsection{Symmetric bidirectional gene flow}

Figure 3 and 4 clearly demonstrate that the method do not systematically show signs of asymmetry 
when gene flow is symmetric. All values, even when sample size and number of loci were low, are close to the expected value of 50\%. When relative migration was calculated from $G_{st}$ (figure 3b and 4b), the result was slightly more variable than when calculated from $D$  (figure 3a and 4a).

\subsubsection{Asymmetric bidirectional gene flow}

The usefulness of the method to detect the underlaying migration pattern, which 
were $1/4$ of the migration rate from population one to population 
two and $3/4$ from population two to population one 
($m*1/4$ and $m*3/4$) is illustrated in figure 5 and 6. The method 
performs best under medium migration rate ($0.005$). 
The correct direction was then estimated in over 80\% of the times for $D$ and $G_{st}$ (figure 5) when the sample size was 20 and over and  
reached 90\% for $G_{st}$ when the sample size exceeded 50 (figure 5b). 
When the number of loci increased, the number of correct directions calculated from $D$ reached over 80\% when the number of loci was 30 or more (figure 6a). For $G_{st}$ the result reached 80\% when the number of loci was 20 and 90\% when the number of loci were 50 or more (figure 6b).

When the migration rate was low ($0.00025$), the number of correct directions for estimates calculated from $D$ was below 75\% for all sample sizes (figure 5a). For $G_{st}$ (figure 5b), 75\% was reached when the sample size reached 40 and stayed around that value as sample size increased further. When the number of loci 
increased, 75\% was reached when the number of loci were 70 for $D$ and 50 for
$G_{st}$, (figure 6) and 80\% was reached for $G_{st}$ 
when the number of loci were 70. 

When the migration rate was high ($0.05$), the method did not estimate the correct direction at a sufficiently high percentage. However the percentage did increased with sample size and number of loci (figure 6 and 7). 

\section{\label{sec:program}Program}

To make the new method described in here accessible and easy for people to use, we have developed a 
web based software application called divMigrate-online.  This user friendly interface, provides 
integrated network visualizations of gene flow patterns among populations, as 
well as, allowing users to test and visualize significant difference in directional 
gene flow between pairs of population samples. This section includes 
a description of the software as well as a demonstration of its use. The flexibility and usefulness  
of the program are demonstrated by analyzing one simulated data set and one published 
microsatellite data set of Atlantic salmon \citep{sandlund14}.

\subsection{\label{sec:aboutprogram} About the program}

divMigrate-online is designed within the shiny (RStudio \& Inc.$2014$) framework for the R 
programming language (Team $2014$). divMigrate-online is written in the R and C++ 
programming languages, integrated using the R package Rcpp \citep{eddelbuettel11} 
and is hosted at (\url{https://popgen.shinyapps.io/divMigrate-online/}).

Estimated gene flow patterns can be visualised and explored using network graphics 
produced using the qgraph R package \citep{epskamp12}. Population samples are 
represented by nodes in the networks. Each node is hypothetically connected to 
every other node by two connections, representing the two reciprocal gene flow 
components between any pair of populations. The properties of these connections, 
such as their length, shading and thickness are determined by the relative strength 
of gene flow. These properties have a particularly beneficial consequence, namely, 
populations that exchange genes at high rates locally, but low rates elsewhere tend 
to cluster together within the network space, thus providing a visual illustration of 
genetic structuring patterns (e.g. figure 8a below). DivMigrate-online also provides 
an useful filter threshold function that makes it possible to show only values above a 
specified number. This is a useful feature when getting to know the data set since it gives the possibility to zoom in and out, especially if many populations are compared simultaneously and the whole picture contains very much information. It can also be useful if low migration rates are 
not of equal interest as high ones.

The divMigrate-online application provides an integrated statistical testing facility, where the asymmetry between migration rates is tested. The statistical significance of differences between directional components of gene flow for a population pair is estimated as follows: 

\begin{enumerate}
\item Resampling of the original data (with replacement) $x$ number of times. 
\item Estimate allele frequencies from resampled data for both populations 
as well as their hypothetical pool of migrants
\item Calculate a user specified relevant parameter (e.g.  $D$ or $G_{st}$ ) between the hypothetical 
pool of migrants and both populations for all $x$ resamples
\item Construct two 95\% confidence intervals of the user specified parameter 
calculated between each population and their shared hypothetical pool of migrants using 
the quantile method (lower probability = 0.025, upper probability = 0.975).
\item Test for overlap of the estimated 95\% confidence intervals. Where there is no overlap, the directional 
gene flow components are said to be significantly different (asymmetric).
\end{enumerate}

This procedure can be coupled with network visualizations, such that only statistically 
significant asymmetric links between population pairs are plotted.

While the web application will be sufficient for most users, there may be those who 
require more flexibility when conducting specialized analyses (e.g. simulation studies 
where batches of files need to be processed). For this purpose the methods provided 
in divMigrate-online are also implemented in the \code{divMigrate} function from the 
\code{diveRsity} R package \citep{Keenan13}. Users interested in using this function can 
find out more by typing  \code{Ô?divMigrateÕ} into the R console (\code{diveRsity} must be installed 
and loaded first). 

\subsection{\label{sec:simprogram} Simulated example}

Using \code{fastsimcoal2} \citep{Excoffier11} five populations connected in a circular stepping stone model with unidirectional
migration
 (1$\rightarrow$2$\rightarrow$3$\rightarrow$4$\rightarrow$5$\rightarrow$1 (rate $0.0005$)) 
was simulated and then analyzed with divMigrate-online. For details about the simulation see 
(\url{https://github.com/lisasundqvist/Sundqvist_et_al_2016}). 

Running the data set in divMigrate-online and choosing $D$ to estimate relative 
migration, the result, as shown in figure 7a is generated. All values except very low ones are visualized in this figure. Depending on the amount of data in the data set divMigrate-online do not always plot all values,  all values are however included and can be seen in a result matrix. The result in figure 7a indicate many migration directions not simulated, this is not surprising given that the simulated migration pattern is circular and that migration between the populations occurred for many generations. Alleles can thus pass through intermediate populations and populations not directly connected can experience indirect gene flow.

To further investigate the data set, we asked divMigrate-online to estimate 
1000 bootstrap iteration, and calculate confidence intervals for all values 
to investigate whether the migration is asymmetric. Figure 7b, 
only shows arrows for the directions that are statistically higher 
relative to the opposite direction. DivMigrate-online 
also provides an useful filter threshold function that makes it possible 
to filter out low values, thus it is easier to visualize 
potentially more relevant patterns. In figure 7c, the filter threshold is set to 0.5 
meaning that only the asymmetric values above 0.5 are shown. 
In this figure, the same migration pattern that was put into the 
simulation is shown. All networks graphics can be exported from the application in various file 
formats, making the creation of publication standard plots straightforward. 
It is also possible to download a result matrix including all numbers.  

\subsection{\label{sec:empprogram} Empirical example}
Recently, \citet{sandlund14} published a study of the spatial and temporal 
genetic structure of the rare landlocked salmon in a fragmented river 
in Norway. Using microsatellite genotype data, the authors identified three distinct genetic clusters. 
The first cluster contained individuals sampled from sites A, B, C and D, 
while the second cluster contained individuals sampled from site E, and the 
third cluster contained individuals sampled from sites F and G. As mentioned previously, 
the network method used to visualize gene flow patterns in divMigrate-online 
has the useful property of representing closely related population samples 
as local clusters within the network space. To demonstrate this the 
contemporary microsatellite data from the original study \citep{datasandlund} 
were re-analyzed by divMigrate-online using Josts $D$. By setting the filter threshold to 0.35 
the genetic structure described in \citep{sandlund14} is reproduced in figure 8a. Indeed, the 
groups shown in figure 8a are similar to the Neighbour-joining 
dendrogram shown in figure 4 of \citep{sandlund14} (excluding the historical 
samples that were not included in the divMigrate-online analysis). Further, we include data of the 
anadromous salmon (ANA) and observe that the landlocked populations are more closely 
related to each other then to the anadromous salmon. In figure 8b, the result is shown with no filter. When
checking for asymmetric gene flow in the system (figure 8c), an interesting result was observed, 
no significant asymmetry was found between the landlocked populations, but all landlocked 
populations exhibited asymmetric gene flow to the anadromous salmon population. Considering the history of this 
system, in particular the fact that these landlocked populations have been isolated from the anadromous salmon for some 9.500 years, it is likely that this pattern is showing us something else. 
In the definition of this method, we use private alleles or alleles only present in one population as a sign of no migration. In a system that historically has been colonized by a small number of individuals, however, it is likely that the colonized 
populations share many of its alleles with its source population. It is also likely that the source population displays a higher level of genetic diversity, and thus are characterized by the presence of  a larger number of  private alleles when compared to the colonized population. It is therefore possible that a strong and recent founder effect is the
cause of what appears to be asymmetrical migration to the method. This is a particular caveat of the method described in here which potential users should be aware of.

\subsection{\label{sec:accessprogram} Accessing the software}
divMigrate-online is a web based software application and can be accessed on any 
operating system through a web browser at \url{https://popgen.shinyapps.io/divMigrate-online/}.
The application can also be launched locally using R. The procedure is as follows:

\begin{enumerate}
\item Install the ÔshinyÕ package in R: \code{install.packages(ÒshinyÓ)}
\item Install \code{diveRsity} from github (instructions can be found at \url{https://github.com/kkeenan02/diveRsity#development-version}
\item Launch divMigrate-online: \code {shiny::runGitHub(Òkkeenan02/divMigrate-onlineÓ)}
\end{enumerate}

The source code for the software is freely available at 

\code {https://github.com/kkeenan02/divMigrate-online}. 
Pull requests can also be made through this repository.

\section{\label{sec:discussion}Discussion}

Here we have introduced a novel alternative approach that estimates directional genetic differentiation and
relative migration from any relevant measure of genetic differentiation, thus
allowing for the detection of asymmetric gene flow patterns.
Tests using simulated data demonstrate that the new approach can 
detect underlying gene flow patterns with high confidence 
when migration is present in either one or two directions. 
The method performed best when gene flow was intermediate ($m = 0.005$). 
In situations where gene flow was high ($m = 0.05$), the method failed to detect underlying 
patterns of migration at a sufficiently high percentage for any of the sample sizes or number of loci combinations tested. 
The low number of times that the simulated direction was found when migration was high is likely linked to the homogenizing effect of high gene flow, resulting in very low differentiation between populations 
(i.e.$F_{st} \approx 0.005$). The directional components of gene flow were thus obscured (i.e. the signal 
to noise ratio was low). However, the upward trend for both increasing sample size and 
number of loci suggest that under suitable conditions the method may be capable of 
resolving such patterns even for cases where genetic structuring is weak.

Based on the simulation results, relative migration calculated from 
$G_{st}$ performed slightly better than those calculated from $D$.
These results are in line with recent observations of the relative properties
of $D$ and $G_{st}$ under variable demographic scenarios \citep{Alcala14}.
Alcala et al. 2014 state that $D$ and $G_{st}$ are complementary measures, and
suggest general guidelines for when the use of $G_{st}$ or $D$ might be 
inappropriate. For example when the expression $\theta < 1< n \theta$ is true, 
(where $\theta$ is the scaled mutation rate ($\theta$ $=2N\mu$) and $n$ is the number of 
populations), as is the case in the simulations used here, the use of 
$G_{st}$ is recommended (table 4 in \citealt{Alcala14}).
Alcala et al., (2014) also introduced a new statistic for the calculation of 
$Nm$ (i.e. the effective number of migrants). This statistic incorporates 
complementary information from both $G_{st}$ and 
$D$, suggesting it may be a more generally suitable measure of migration. When using $Nm_{Alcala}$ for calculating the percent of correct directions in the different simulation scenarios, the result is very similar to the result obtained when using $G_{st}$. The simulation results for $Nm_{Alcala}$ is shown in Appendix 1. 
In divMigrate-online it is possible to calculate directional relative migration from $Nm_{Alcala}$
as well as from $D$ and $G_{st}$.

The network plots in divMigrate-online has the very useful property of representing similar population samples as local clusters within the network space. 
Because the directional migration calculations require that populations are pre-determined, 
this feature is likely to be useful only as a confirmatory complement to more quantitative 
methods such as those implemented in STRUCTURE \citep{pritchard00}, 
ADEGENET \citep{jombart08} or DAPC \citep{jombart10}. The filter threshold function makes 
it easy to visualize a data set, but how to filter a data set is subjective and it is therefore important 
to clearly state when and how this function is used.

In certain systems estimates of directional relative migration may be influenced by historic demography, as discussed in the empirical example with the landlocked Salmon. Where historical events such as founder effects or recent common ancestry have influenced the genetic composition of individuals and/or populations, footprints will be left. These footprints may enhance or diminish signs of recent migration in the genetic data. The information about migration is then obscured and makes it difficult for any method to correctly describe the migration patterns correctly. The result of the method can never reveal more information than what is to be found in the allele frequencies examined. Future work might include a more comprehensive exploration of how good the method is to detect signs of founder effects and if it is possible to distinguish them from asymmetric migration. It would also be interesting to investigate how recent common ancestry affects the ability of the method to detect contemporary migration patterns. The composition of populations may also lead to complications. In small populations fore instance genetic drift plays a bigger role. If a migrating allele is lost due to genetic drift the trace of that migration event will also be lost. If populations are uneven in size the relative impact of an allele will be different in the different populations and it will be difficult to interpret the meaning of the directional relative migration calculated, especially if one of the populations is small. By weighing the allele frequencies of populations proportionally to local size or reproductive values one could probably improve the results of the method in such situations \citep{Hossjer14}. In addition, it remains to be examined how robust the method is when applied to system not at equilibrium \citep{Boileau1992} and with influences of ghost populations \citep{beerli04}. 

When using allele frequency data, it is possible that alleles present at low frequencies are 
underestimated due to sampling effects \citep{fung2014, Gautier2013}.
Since less common alleles (or alleles only present in one population) are used as 
an indication of no gene flow, rates estimated by the model may be slightly 
inflated if applied to imprecisely estimated allele frequencies.
However, using alleles only present in one population as an indication of gene flow is a common strategy.
Slatkin called these \emph{"private alleles''} and showed that the logarithmic 
average frequency of private alleles are approximately linearly related to the 
logarithm of $Nm$ \citep{Slatkin85}. This issue is somewhat dealt with in our model, 
since its focus is only on relative migration, thus, making it possible to assume
that the probability of underestimating low frequency alleles is equally high in all 
populations.

Other available approaches for calculating asymmetric migration, such as 
Migrate \citep{Beerli09} and BayesAss \citep{Wilson03}, while powerful, 
are known to be difficult to use correctly, due, in part, to the large 
number of parameters and options that need to be adjusted to the data set 
under consideration. Thus, the use of these programs as black-boxes can 
lead to misleading results assigned a high confidence \citep{Faubet07}.
Furthermore, these programs are also computationally demanding and, hence, can sometimes 
take impractical amounts of time to run. In comparison, the method presented here is 
easy to use, conceptually tangible and computationally efficient.

In conclusion, by introducing the concept of directional genetic 
differentiation, the novel method presented  in here, enables users to gain new knowledge 
about genetic structure in systems that experience asymmetric gene flow.
By acquiring such information, it is possible to examine the relationships between the direction of migration
to correlated factors such as wind or water currents, which has been done in \citet{Godhe2013}, \citet{Sjoqvist2015} and \citet{godhe15}.
This can lead to new insights about evolutionary processes, as well as allowing for more 
accurate predictions for the purposes of conservation and management.
Of particular note is the utility of the method for understanding 
meta-populations and their source-sink dynamics. More specifically, the 
ability to identify low quality sink populations as well as high quality 
source populations is a major challenge in conservation genetics, and is 
readily possible using the method presented here.


\section{\label{sec:conclusions}Acknowledgements}
We kindly acknowledge Per Jonsson for valuable comments on the manuscript and
Anna Godhe and Kerstin Johannesson for valuable discussions. We also acknowledge helpful reviewers for valuable comments that considerably improved the paper. We are also grateful to the authors of Sandlund et al, (2014) for making their data publicly and freely available. LS was supported by the Faculty of Science at the University of Gothenburg. DK was supported by the Linnaeus Centre for Marine Evolutionary Biology at the University of Gothenburg (www.cemeb.science.gu.se) founded by a Linnaeus-grant from the Swedish Research Councils, VR and Formas. KK was supported by a Ph.D studentship from the Beaufort Marine Research Award in Fish Population Genetics, funded by the Irish Government under the Sea Change programme, the National Marine Knowledge, Research and Innovation Strategy 2007-2013. PAP was also supported by this award. KK also received support by way of a Research Award from the Department of Culture, Arts and Leisure, Northern Ireland (R3579BSC).

\section{\label{sec:acc}Data Accessibility} 
Only simulated and already published data is used in this manuscript, computer code is freely available at github repository (\url{https://github.com/lisasundqvist/Sundqvist_et_al_2016}).

\bibliographystyle{apalike}

\clearpage
\section{\label{sec:figures} Figures and tables}
\begin{figure}[htbp]
\centering
\includegraphics[width=\textwidth]{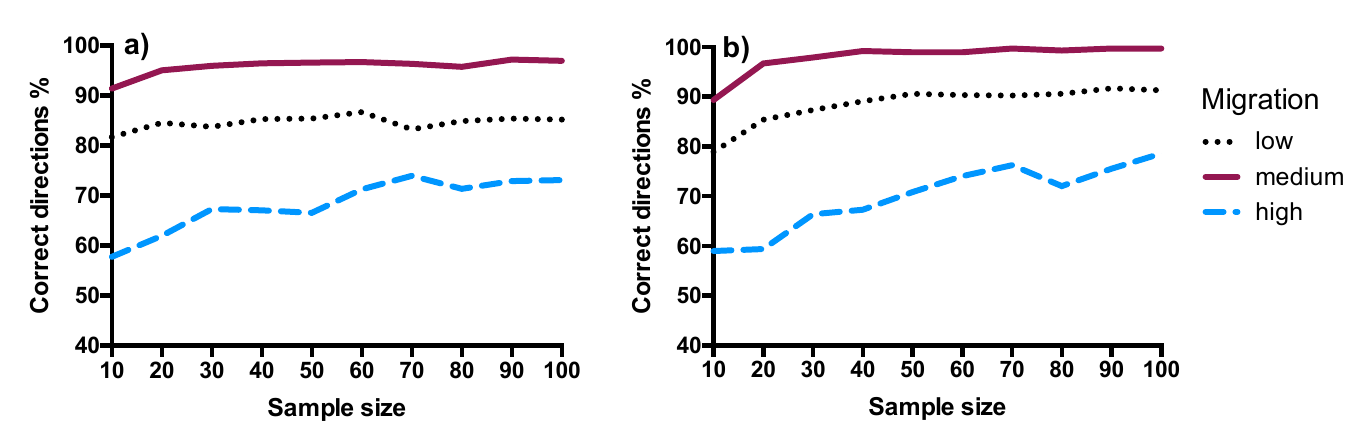}
 \caption{Unidirectional migration: percent correct directions as a function of sample size calculated using $D$ (a) and $G_{st}$ (b).
   Increasing sample size is evaluated at high ($0.05$), medium ($0.005$) and low ($0.00025$) gene flow. The number of loci are kept fixed at 50. }
\label{fig:unidirectionN}
\end{figure}

\begin{figure}[htbp]
\centering
\includegraphics[width=\textwidth]{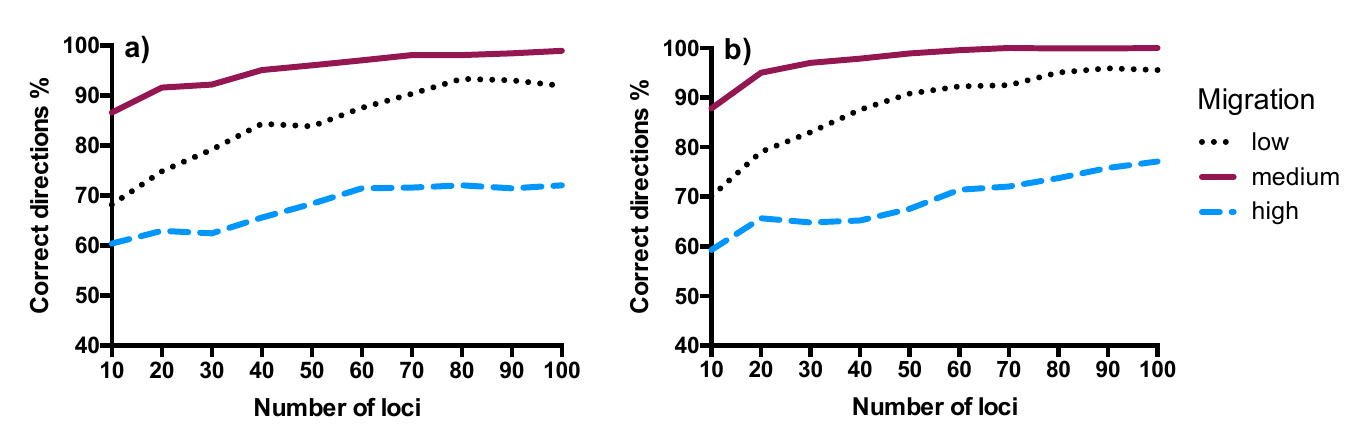}
  \caption{Unidirectional migration: percent correct directions as a function of number of loci calculated using $D$ (a) and $G_{st}$ (b).
   Increasing number of loci are evaluated at high ($0.05$), medium ($0.005$) and low ($0.00025$) gene flow. The sample size is kept fixed at 50. }
\label{fig:unidirectionlocus}
\end{figure}

\begin{figure}[htbp]
\centering
\includegraphics[width=\textwidth]{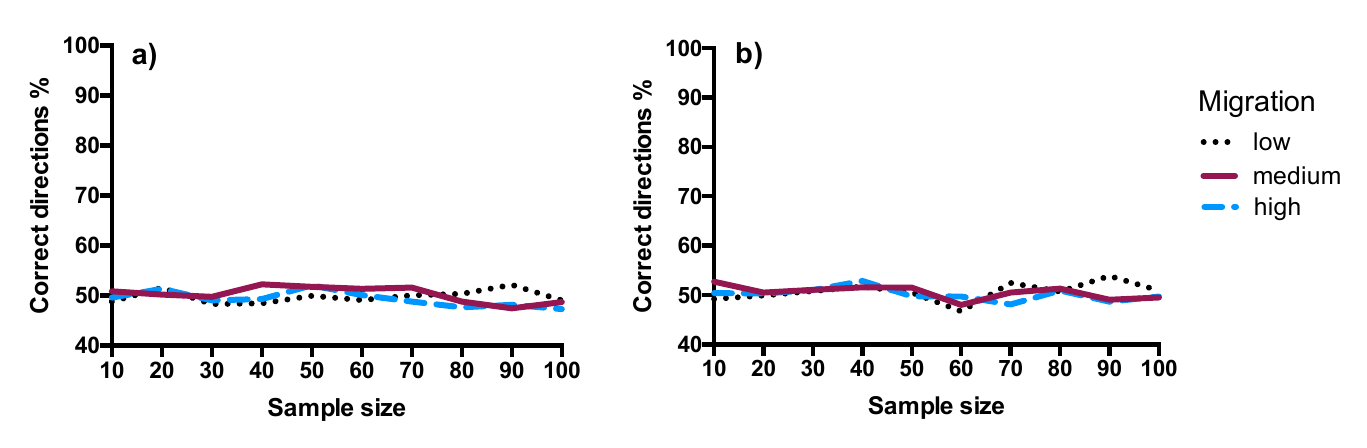}
   \caption{Bidirectional symmetric migration: percent correct directions as a function of sample size calculated using $D$ (a) and $G_{st}$ (b).
   Increasing sample size is evaluated at high ($0.05$), medium ($0.005$) and low ($0.00025$) gene flow. The number of loci are kept fixed at 50. When migration is symmetric the expected value is 50\%.}
\label{fig:bidirectionN}
\end{figure}

\begin{figure}[htbp]
\centering
\includegraphics[width=\textwidth]{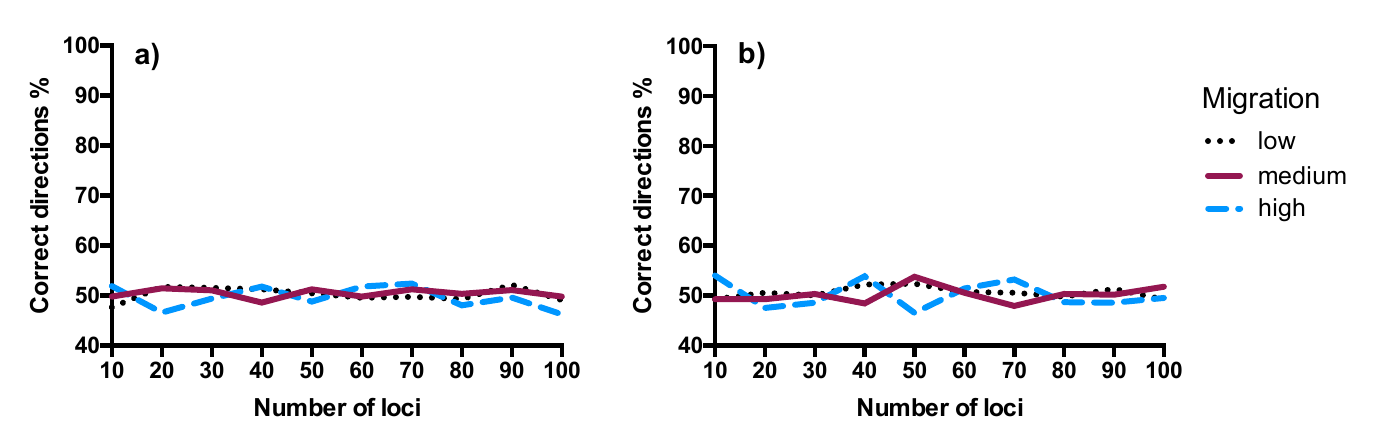}
   \caption{Bidirectional symmetric migration: percent correct directions as a function of number of loci calculated using $D$ (a) and $G_{st}$ (b).
   Increasing number of loci are evaluated at high ($0.05$), medium ($0.005$) and low ($0.00025$) gene flow. The sample size is kept fixed at 50. When migration is symmetric the expected value is 50\%.}
\label{fig:bidirectionlocus}
\end{figure}

\begin{figure}[htbp]
\centering
\includegraphics[width=\textwidth]{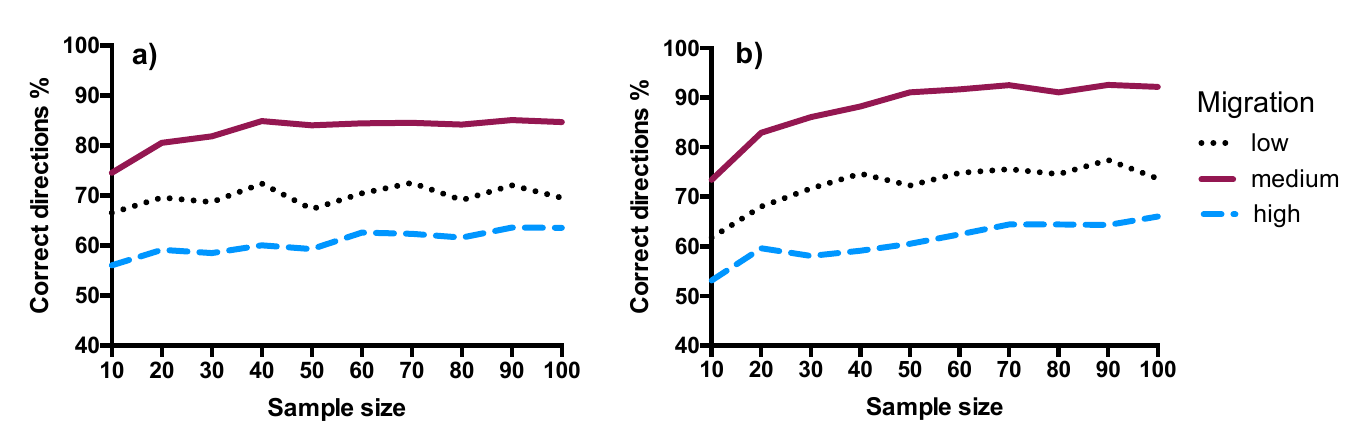}
   \caption{ Bidirectional asymmetric migration: percent correct directions as a function of sample size calculated using $D$ (a) and $G_{st}$ (b).
   Increasing sample size is evaluated at high ($0.05$), medium ($0.005$) and low ($0.00025$) gene flow. The number of loci are kept fixed at 50. }
\label{fig:biasymN}
\end{figure}

\begin{figure}[htbp]
\centering
\includegraphics[width=\textwidth]{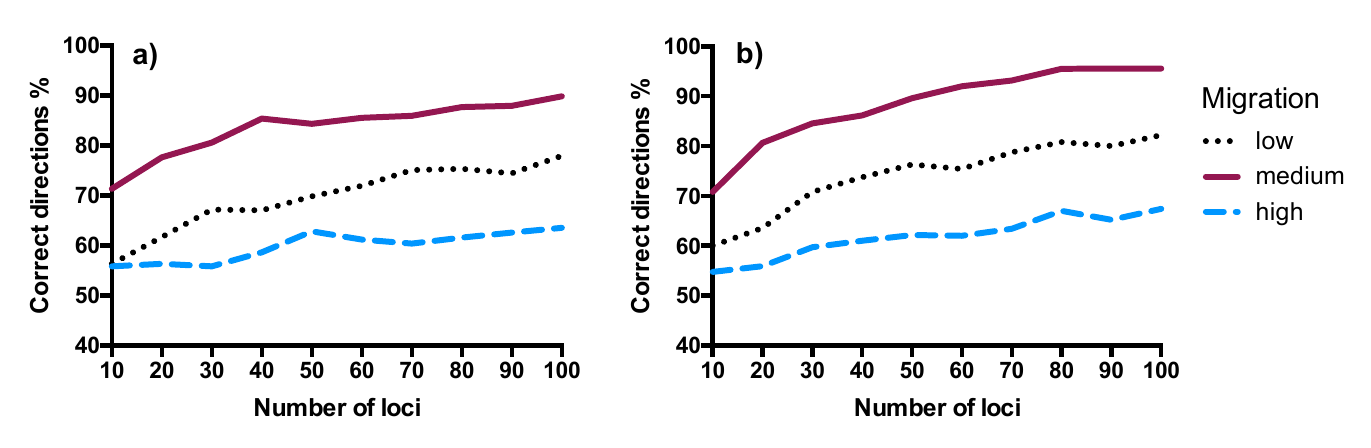}
   \caption{Bidirectional asymmetric migration: percent correct directions as a function of number of loci calculated using $D$ (a) and $G_{st}$ (b).
   Increasing number of loci are evaluated at high ($0.05$), medium ($0.005$) and low ($0.00025$) gene flow. The sample size is kept fixed at 50. }
\label{fig:biasymNloci}
\end{figure}

\begin{figure}[htbp]
\centering
\includegraphics[width=\textwidth]{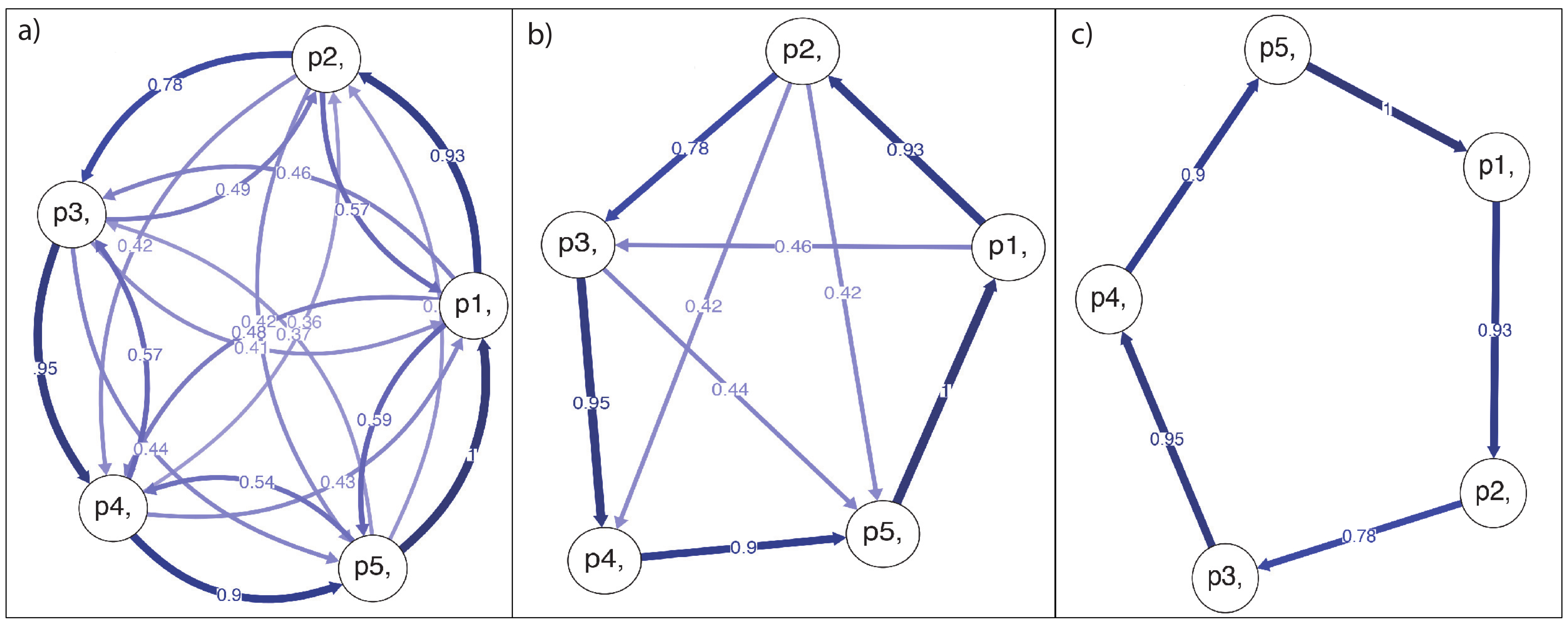}
   \caption{ Directional relative migration calculated by divMigrate-online for the simulated circular stepping stone model with unidirectional
migration. Figure a) illustrate the calculated migration values. Figure b) only include the values found to be asymmetric, i.e. they are statistically higher in the shown direction. In figure c) the filter threshold for the asymmetric values are set to 0.5.}
\label{fig:allinone}
\end{figure}

\begin{figure}[htbp]
\centering
\includegraphics[width=\textwidth]{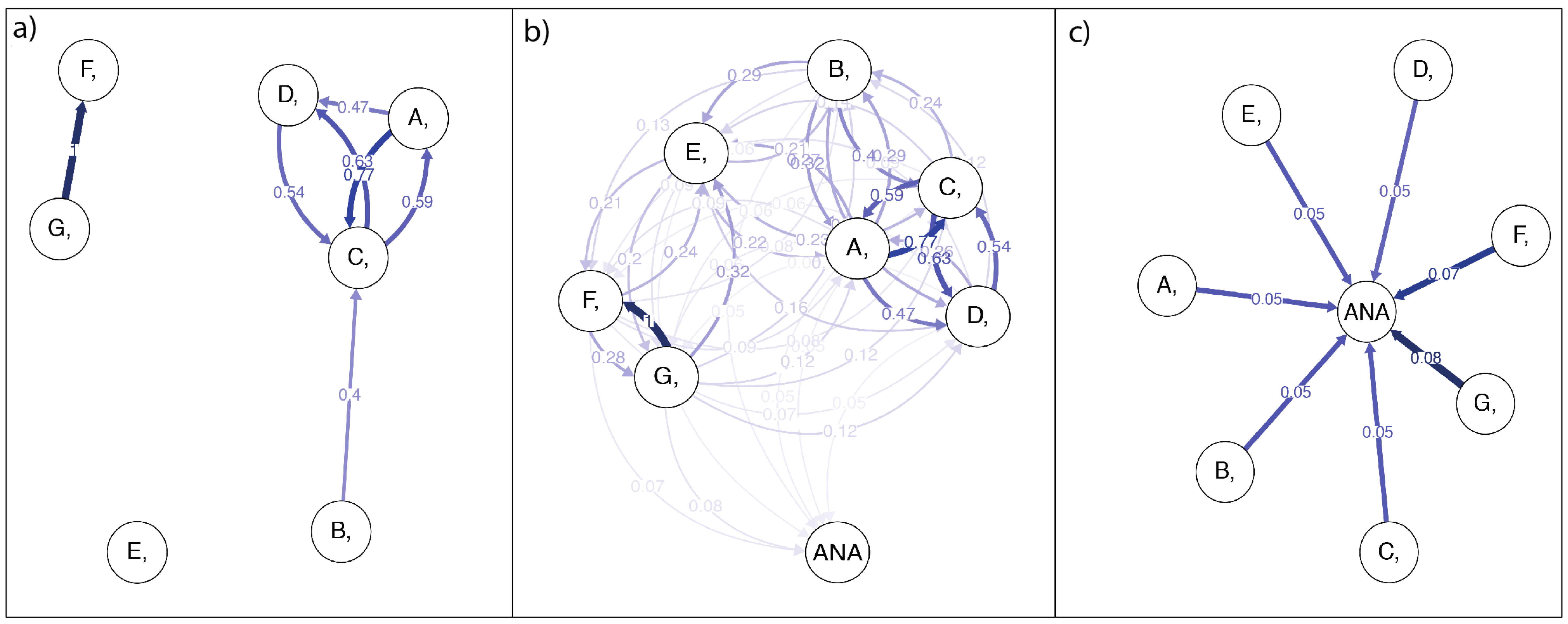}
   \caption{Directional relative migration estimated by divMigrate-online. Figure a) includes population A to G and the filter threshold is set to 0.35. In figure b) data from the anadromous salmon (ANA) is included, no filter is used. Figure c) illustrate the migration that is found to be asymmetric, no filter is used.}
\label{fig:allinonesalmor}
\end{figure}

\begin{table}[htbp]
  \begin{tabular}{|l||l|l|}
    \hline&\textbf{Population A}&\textbf{Population B}\\\hline\hline
    \textbf{Allele 1}&$0.4$&$0.2$\\\hline
    \textbf{Allele 2}&$0.6$&$0.3$\\\hline
    \textbf{Allele 3}&$0.0$&$0.5$\\\hline
  \end{tabular}¥
  \caption{\label{tab:thought-exp}Allelic matrix $A$ of the thought
    experiment consisting of two populations A and B with directional
    gene flow from A to B.}
\end{table}

\clearpage
\section{\label{sec:appendix}Appendix}
\renewcommand{\thefigure}{A\arabic{figure}}

\setcounter{figure}{0}

\begin{figure}[htbp]
\centering
\includegraphics[width=\textwidth]{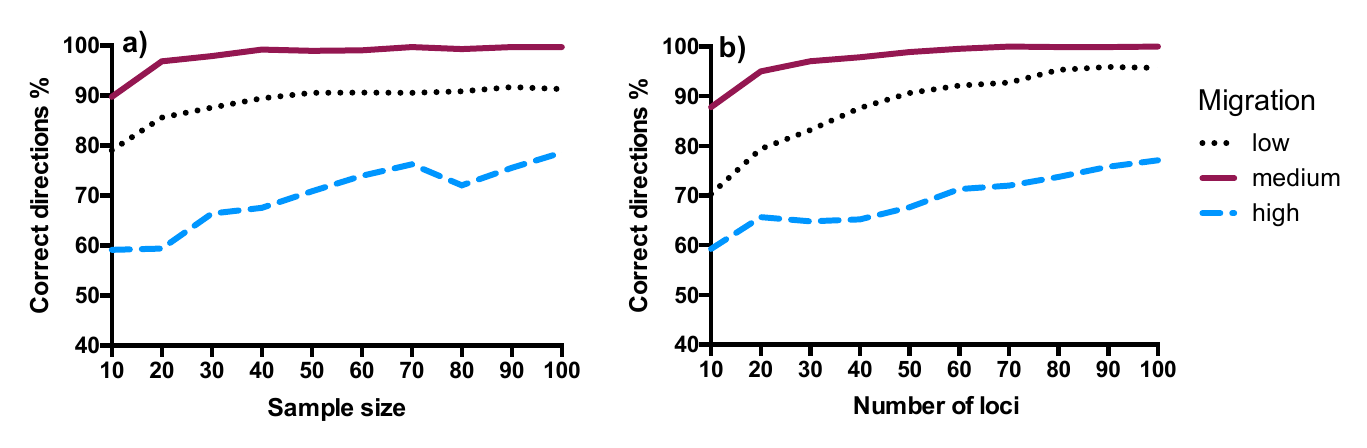}
\caption{Unidirectional migration: percent correct directions as a function of sample size (a) and number of loci (b) calculated using $Nm_{Alcala}$. Increasing sample size and number of loci are evaluated at high ($0.05$), medium ($0.005$) and low ($0.00025$) gene flow. When sample size is evaluated the number of loci are kept fixed at 50 and when the number of loci are evaluated the sample size is kept fixed at 50.}
\label{fig:uninm}
\end{figure}

\begin{figure}[htbp]
\centering
\includegraphics[width=\textwidth]{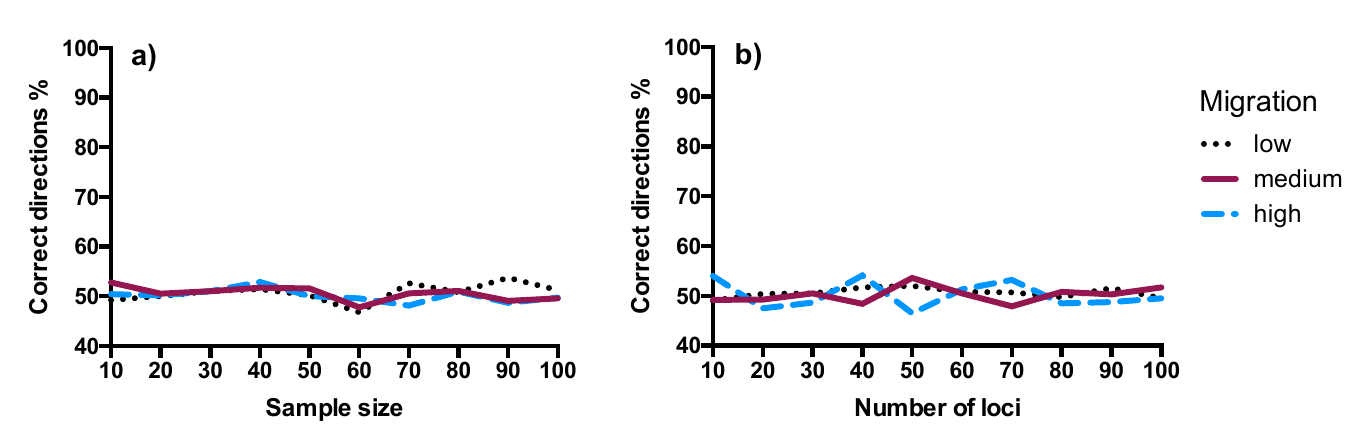}
\caption{Bidirectional symmetric migration: percent correct directions as a function of sample size (a) and number of loci (b) calculated using $Nm_{Alcala}$. Increasing sample size and number of loci are evaluated at high ($0.05$), medium ($0.005$) and low ($0.00025$) gene flow. When sample size is evaluated the number of loci are kept fixed at 50 and when the number of loci are evaluated the sample size is kept fixed at 50. When migration is symmetric the expected value is 50\%.}
\label{fig:symnm}
\end{figure}

\begin{figure}[htbp]
\centering
\includegraphics[width=\textwidth]{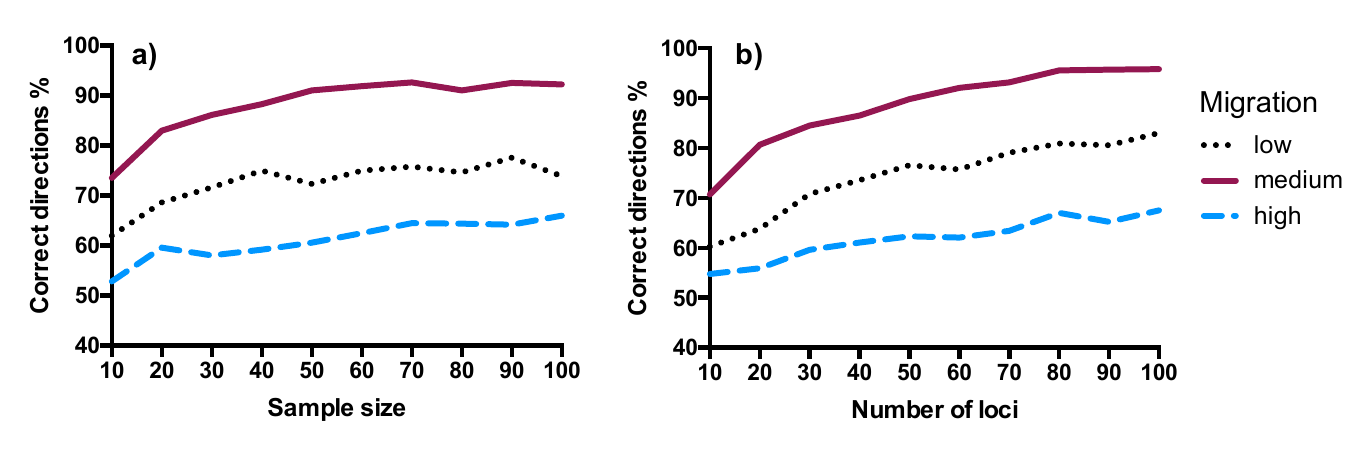}
\caption{Bidirectional asymmetric migration: percent correct directions as a function of sample size (a) and number of loci (b) calculated using $Nm_{Alcala}$. Increasing sample size and number of loci are evaluated at high ($0.05$), medium ($0.005$) and low ($0.00025$) gene flow. When sample size is evaluated the number of loci are kept fixed at 50 and when the number of loci are evaluated the sample size is kept fixed at 50.}
\label{fig:asymnm}
\end{figure}

\end{document}